\documentclass[a4paper,10pt,twoside]{cpc-hepnp}
\usepackage{CJK,upgreek,fancyhdr}
\usepackage{multicol}
\usepackage{graphicx}
\usepackage{booktabs}
\usepackage{amssymb,bm,mathrsfs,bbm,amscd}
\usepackage[tbtags]{amsmath}
\usepackage{lastpage}
\usepackage{color}
\usepackage{caption}
\usepackage{subfigure}

\begin{document}
\begin{CJK*}{GB}{gbsn}

\fancyhead[c]{\small Chinese Physics C~~~Vol. xx, No. x (201x) xxxxxx}
\fancyfoot[C]{\small 010201-\thepage}

\footnotetext[0]{Received 6 April 2017}

\title{The effect of cavity length detuning on the output characteristics for the middle infrared FEL oscillator of FELiChEM\thanks{Supported by National Natural Science
Foundation of China (No. 21327901, 11205156) }}

\author{%
	Zhou-Yu Zhao(ÕÔÖÜÓî)
	\quad He-Ting Li(ÀîºÍÍ¢)$^{1)}$\email{liheting@ustc.edu.cn}%
	\quad Qi-Ka Jia(¼ÖÆô¿¨)
}
\maketitle

\address{%
National Synchrotron Radiation Laboratory, University of Science and Technology of China, Hefei, 230029, Anhui, China\\
}

\begin{abstract}
FELiChEM is an infrared free electron laser (FEL) facility under construction, which consists of two oscillators generating middle-infrared and far-infrared laser covering the spectral range of 2.5-200 ${\rm{\mu m}}$. In this paper, we numerically study the output characteristics of the middle-infrared oscillator with the accurate cavity length detuning. Emphasis is put on the temporal structure of the micropulse and the corresponding spectral bandwidth. Taking the radiation wavelengths of 50 ${\rm{\mu m}}$ and 5 ${\rm{\mu m}}$ as examples, we show that the output pulse duration can be tuned in the range of 1-6 ps with corresponding bandwidth of 13-0.2\% by adjusting the cavity length detuning. In addition, a special discussion on the comb structure is presented, and it is indicated that the comb structure may arise in the output optical pulse when the normalized slippage length is much smaller than unity. This work has reference value for the operation of FELiChEM and other FEL oscillators.
\end{abstract}



\footnotetext[0]{\hspace*{-3mm}\raisebox{0.3ex}{$\scriptstyle\copyright$}2013
Chinese Physical Society and the Institute of High Energy Physics
of the Chinese Academy of Sciences and the Institute
of Modern Physics of the Chinese Academy of Sciences and IOP Publishing Ltd}%

\begin{multicols}{2}

\section{Introduction}

  Free electron lasers (FELs) hold great promise as a high-power light source with a tunable wavelength. Many FEL user facilities have been constructed and are being proposed worldwide, from the far infrared to hard x-ray spectral region \cite{lab1,lab2,lab3,lab4,lab5}. These facilities are based on different kinds of FEL schemes according to their object radiation wavelength and the demand of output properties, e.g., self-amplified spontaneous emission for hard x-ray FEL \cite{lab1,lab6} and seeded FEL for the extreme ultraviolet FEL \cite{lab2,lab7}. In the infrared (IR) and terahertz (THz) region, oscillator FEL is one of the most important FEL schemes. Nowadays infrared FEL oscillators are built worldwide as user facilities, such as CLIO in France \cite{lab4,lab8}, FHI-FEL in Germany \cite{lab9,lab10}, FELIX in the Netherlands \cite{lab11,lab12} and so on. The recently proposed scheme using the pre-bunched electron bunch train provides a way to build a compact infrared and THz source, however, its optimized working frequency region is in the range of 1-5 THz \cite{lab13,lab14}.

Recently, a new FEL user facility named FELiChEM is under construction in China, which consists of two oscillators driven by one RF Linac, and will be used to generate middle-infrared (2.5-50 $\rm{\mu}$m) and far-infrared (40-200 $\rm{\mu}$m) laser \cite{lab15}. It will be a dedicated IR light source aiming at energy chemistry research and the first light is targeted for the end of 2017. The FEL output characteristics are influenced by many factors. In an FEL oscillator (FELO), detuning the cavity length is an important method to control the output characteristics of the radiation and obtain the object pulse energy, pulse length and bandwidth \cite{lab16,lab17}. Considering this point, the optical cavity of FELiChEM was designed with the function of convenient and accurate cavity length detuning.

In this paper, we numerically study the output characteristics for FELiChEM with the cavity length detuning. Firstly, we briefly review the relative theory in Section 2. Then in Section 3, we simply describe the FELiChEM parameters and numerically investigate the optical output characteristics with the cavity length detuning. A special focus is put on the comb structure inside the optical pulse in Section 4. Finally, we summarize in the last section.

\section{Brief review of relevant theory}
In an FELO the optical pulse and electron beam are coupled as electrons pass through the undulator. The optical pulse goes back and forth inside the cavity and overlaps with fresh electron bunch in each round trip. The radiation is reinforced until gain equals to the total cavity loss. However, because of the different velocities between electron beam and optical pulse, the optical pulse gradually advances the electrons. The slippage length is
  \begin{eqnarray}
\Delta S = N_u \lambda _s
\end{eqnarray}
where $N_u$ is the period number of undulator, $\lambda _s $ is the resonant wavelength. The slippage effect induces a longitudinal mismatch, which will be reinforced in each round trip and reduce the saturated power. Therefore, the cavity length detuning $\Delta L$ is introduced to improve the coupling efficiency. It should be noticed that the zero detuning ($\Delta L =0$) is defined as the cold cavity length required to obtain the perfect overlap between the optical pulse and the electron bunch, and a positive value of $\Delta L$ means the shortening with respect to the ideal cavity length. According to super-mode theory \cite{lab18}, the normalized slippage length $\mu$ is defined as
\begin{eqnarray}
\mu  = \frac{{\Delta S}}{{\sigma _e }}
\end{eqnarray}
where $\sigma _e$ is RMS electron bunch length. The optical field inside the cavity can be considered as consisting of many super modes. The gain of these modes specifies the FEL gain. Without considering the 3D effects, when the small signal gain $g_0$ is not too large, the total FEL gain $G(g_0, \mu, \theta)$ and optical power $W_p (g_0, \mu, \theta, \eta)$ can be simplified as \cite{lab19}.
\begin{eqnarray}
G({g_0},\mu ,\theta ) \approx 1.86{g_0}\theta [1 - \ln (2.19\theta  + 0.72\mu \theta )]
\end{eqnarray}
\begin{eqnarray}
 {{W}_{p}}({{g}_{0}},\mu ,\theta ,\eta )=2.414[\sqrt{\frac{{{\theta }_{0}}}{\theta (1+0.33\mu )}}\times \nonumber  \\
 \exp [0.5(1-\frac{\eta {{\theta }_{0}}}{0.85{{g}_{0}}\theta (1-\eta )})]-1]
\end{eqnarray}
where $\theta _0 =0.456$, $\eta$ is the cavity loss, and $\theta={{ {4\Delta L} } \mathord{\left/
 {\vphantom {{ {\Delta L} } {l_e g_0 }}}\right.
 \kern-\nulldelimiterspace} {(g_0 \Delta S) }} $ is the cavity detuning parameter.

Furthermore, if $\mu$ is larger than $1$, the optical pulse width $\sigma _p$ is approximately proportional to $({{\left| {\Delta L} \right|} \mathord{\left/
 {\vphantom {{\left| {\Delta L} \right|} {l_e g_0 }}} \right.
 \kern-\nulldelimiterspace} {\sigma _e g_0 }})^{{1 \mathord{\left/
 {\vphantom {1 3}} \right.
 \kern-\nulldelimiterspace} 3}} $. It is obvious that $\sigma _p$ becomes shorter with the decrease of $\Delta L$. $\Delta L =0$ corresponds to the minimum optical pulse length in case of the laser lethargy effect being ignored. However, when laser lethargy is included, the minimum pulse length occurs when $\Delta L >0$. In other words, the electron bunch and optical pulse are not in perfect synchronism.

On the other hand, if $\mu$ is less than $1$, due to the trapped-particle instability, a series of sub-pulses are probably generated inside the optical pulse which is the so-called comb structure \cite{lab20}. The dependence of $\sigma _p$ on $\Delta L$ becomes complicated. Generally speaking, when the cavity detuning is not too large, the width of one single spike is comparable to the coherent length $\l _c$ and $\sigma _p$ is comparable to $\sigma _e$ in deep saturation region. The coherent length $\l _c$ is defined as
\begin{eqnarray}
\l _c  = \frac{{\lambda _s}}{{4\pi\rho }}
\end{eqnarray}
where $\rho$ is the FEL Pierce parameter \cite{lab21}.

With the discussion above, one can realize that $\mu$ and $\theta$ have a strong impact on the longitudinal pulse profile, while peak power can be tuned by varying $\theta$. Moreover, the comb structure may be observed when $\mu$ is less than one, which is associated with the trapped-particle instability, and as such have a duration that is set by the building process in the radiation field. In addition, when the small signal gain $g_0$ is larger than 0.3, the equation (3) and (4) should be modified, which has been detailedly discussed in Ref. \cite{lab19}.

\begin{center}
	\tabcaption{ \label{tab1}  Parameters for the wavelengths of 5 $\mu$m and 50 $\mu$m of Mid-infrared FELO of FELiChEM.}
	\footnotesize
	\begin{tabular*}{80mm}{c@{\extracolsep{\fill}}ccc}
		\toprule Parameter  & Specification & Unit \\
		\hline
        radiation wavelength & 5, 50\hphantom{0}& $\mu$m  \\		
        beam energy\hphantom{00}  & 50, 25\hphantom{0} & MeV \\
		beam transverse emittance\hphantom{00} &30\hphantom{0}& $\mu$m.rad   \\
		beam energy spread & 0.5\%, 1\%\hphantom{0}& -   \\
		rms bunch length & 4\hphantom{0}& ps  \\
		bunch charge & 1\hphantom{0}& nC  \\
		undulator type & Planar\hphantom{0}& -  \\
		undulator period & 4.6\hphantom{0}& cm  \\
        undulator parameter $K$ & 1.47, 2.89\hphantom{0}& -   \\
		period number & 50\hphantom{0}& -  \\
		cavity length & 5.04\hphantom{0}& m  \\
        rayleigh length & 0.77\hphantom{0}& m  \\
        outcoupling hole diameter & 1.0, 3.0\hphantom{0}& mm  \\
		\bottomrule
	\end{tabular*}
\end{center}

\begin{center}
\includegraphics[height=4cm,width=8cm]{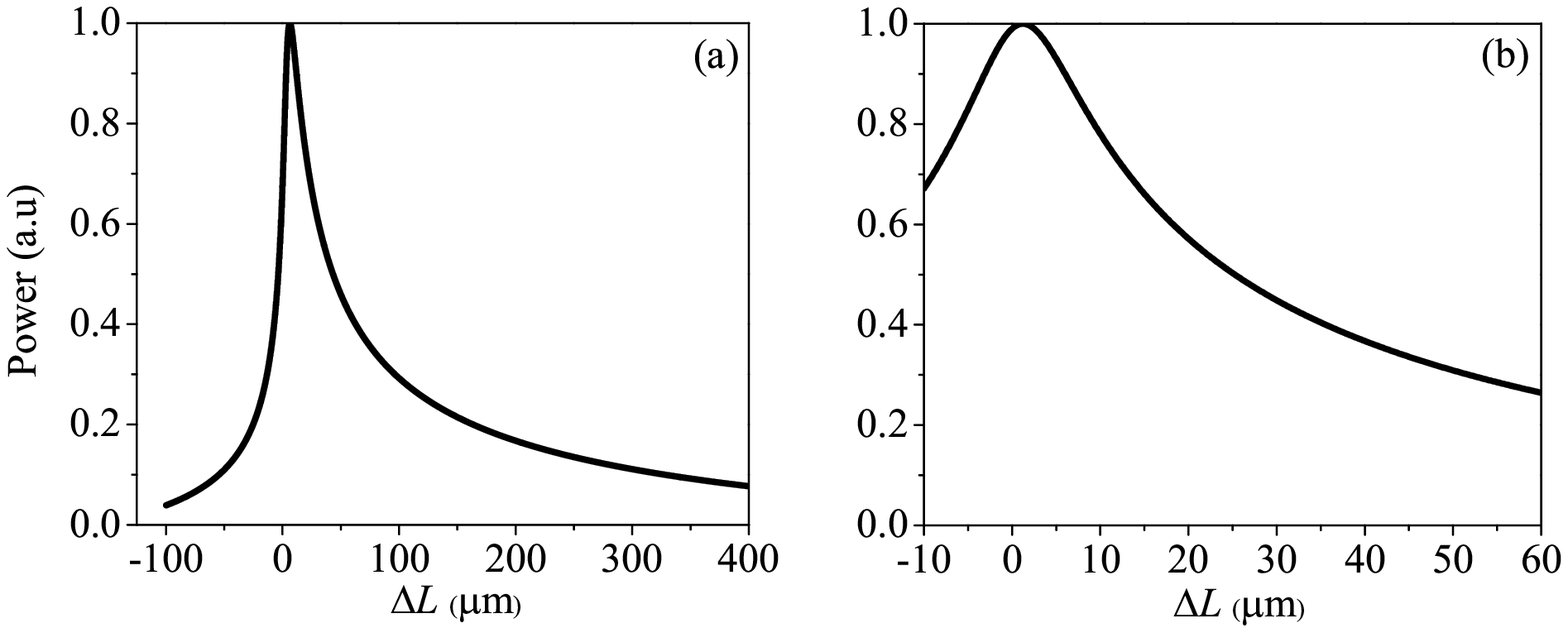}
\figcaption{\label{fig1}   The 1-D detuning curves of the cavity length for the radiation wavelengths of (a) 50 $\mu$m and (b) 5$\mu$m.}
\end{center}
\begin{center}
	\includegraphics[height=4cm,width=8cm]{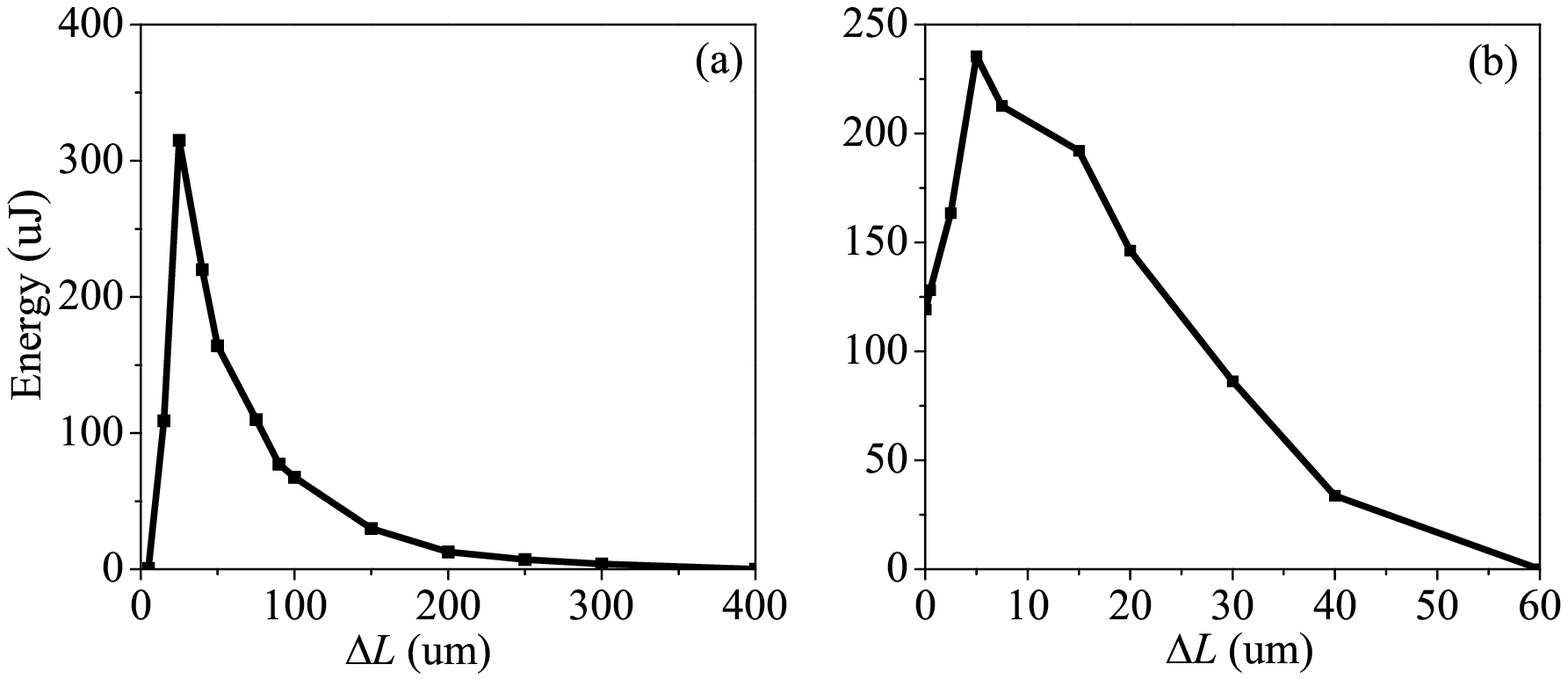}
	\figcaption{\label{fig2}   The 3-D detuning curves of the cavity length for the radiation wavelengths of (a) 50 $\mu$m and (b) 5 $\mu$m. }
\end{center}

\section{Simulations}
The middle infrared FELO of FELiChEM covers the spectral range of 2.5-50 $\mu$m, and here we take the wavelengths of 5 $\mu$m and 50 $\mu$m as examples for simulations. The corresponding parameters are listed in Table 1. For 50 $\mu$m and 5 $\mu$m cases, the small signal gains $g_0$ are 0.46 and 0.96, respectively, and the out-coupling rates are calculated to be 5\% and 6\%, respectively. Based on the super-mode theory, the 1-D detuning curves are given in Fig. 1, which show that the output powers for both the two wavelengths achieve their maximum values when the detuning is slightly larger than zero, due to that the optical pulse travels slower than the speed of light in vacuum. From Fig .1, the optimal detunings for maximizing the output power are 6.5 $\mu$m and 1.2 $\mu$m, respectively.

Actually, many other factors also have influences on the output properties, such as the electron beam emittance and so on. Therefore, we numerically study the output characteristics by three-dimension simulations with GENESIS code \cite{lab22} in combination with OPC code \cite{lab23}. However, the GENESIS code limits the cavity length detuning only to half-integer multiples of radiation wavelength. With a external script that shifts the temporal position of the electron beam by an
arbitrary step with respect to the FEL pulse in each round trip, we are able to detune the cavity length finely.

The cavity detuning curves for FEL pulse energy based on 3-D simulation are shown in Fig. 2. The curves have almost same shape compared with those depicted in Fig. 1. The optimal detuning length $\Delta {{L}_{op}}$ is defined by the maximum energy. Thus $\Delta {{L}_{op}}$ for the two wavelengths are 25 $\mu$m and 5 $\mu$m respectively, which are approximately 4 times bigger than those from 1-D analysis. Actually, it should be based on the experiments to decide the specific detuning then acquiring the optimal output characteristics.

With the optimal cavity length detuning, the evolution of output pulse energy is shown in Fig. 3. The pulse energies increase rapidly in the first 50 round trips and  saturate at 320 $\mu$J and 240 $\mu$J, respectively.

\begin{center}
\includegraphics[height=3.8cm,width=8cm]{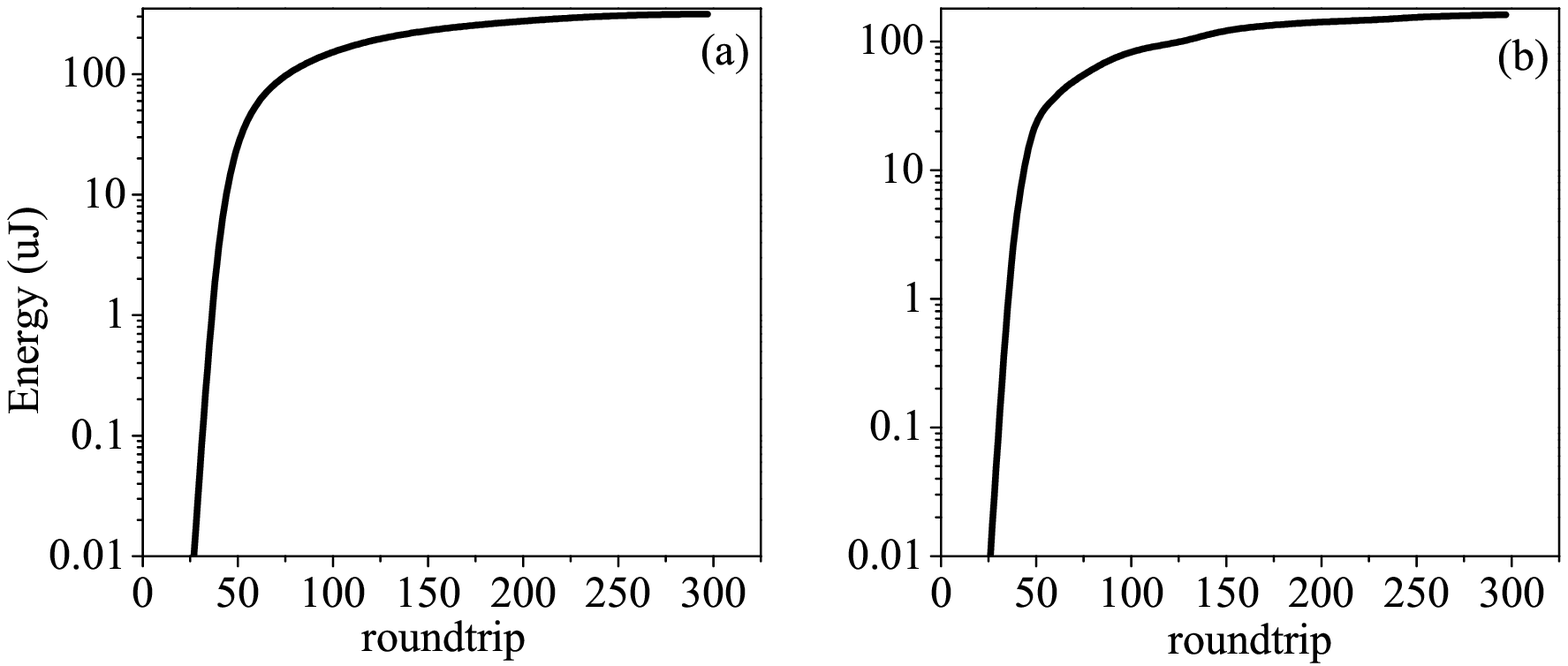}
\figcaption{\label{fig3}   The evolution of micropulse energy for the radiation wavelengths of (a) 50 $\mu$m and (b) 5 $\mu$m with the optimal cavity length detuning of 25 $\mu$m, 5 $\mu$m, respectively.}
\end{center}

For the wavelength of 50 $\mu$m case, Fig. 4 shows the corresponding temporal and spectral structures of the optical pulse at saturation with different cavity length detuning . The results indicate that:
1) When $\Delta L$$<$25  $\mu$m, the pulse width decreases and the spectral bandwidth increases with the increase of $\Delta L$;
2) When $\Delta L$$>$25  $\mu$m, the pulse width increases and the spectral bandwidth decreases with the increase of $\Delta L$.

In our simulations, the optimal detuning $\Delta {{L}_{op}}$ corresponds to the narrowest pulse width and the widest spectral bandwidth, which is consistent with the analysis of super-mode theory in Section 2 and was firstly proposed in the study of Ref. \cite{lab19}. From these results, one can see that the pulse width can be tuned in the range of 1-3.2 ps with corresponding RMS bandwidth 13-1\% by adjusting $\Delta L$ from 5 $\mu$m to 100 $\mu$m.
\end{multicols}
\begin{center}
\includegraphics[width=14cm]{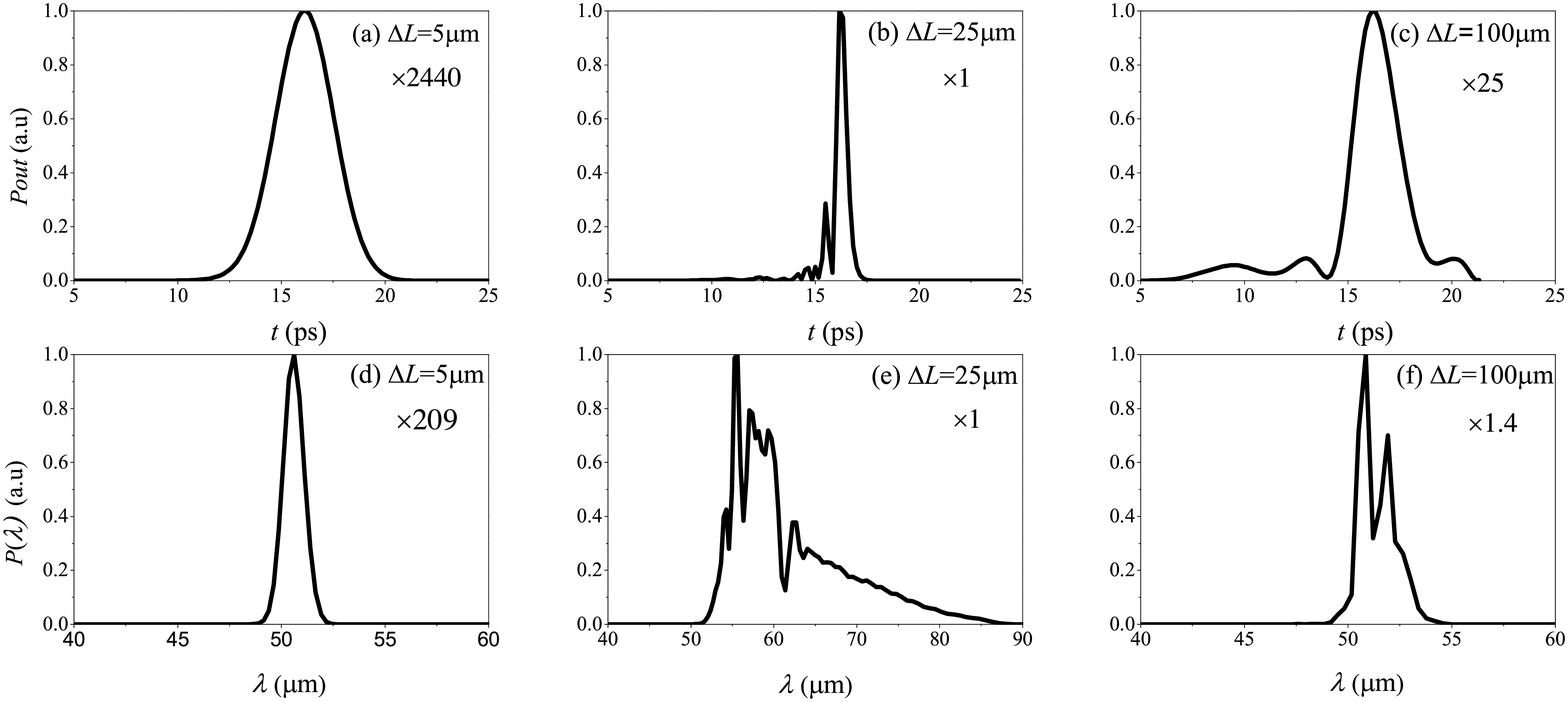}
\figcaption{\label{fig4} The temporal and spectral structures of the FEL pulses at saturation with different detuning for the wavelength of 50 $\mu$m. Both the intensity values are normalized to those of $\Delta L=25$ $\mu$m.}
\end{center}

\begin{center}
\includegraphics[width=14cm]{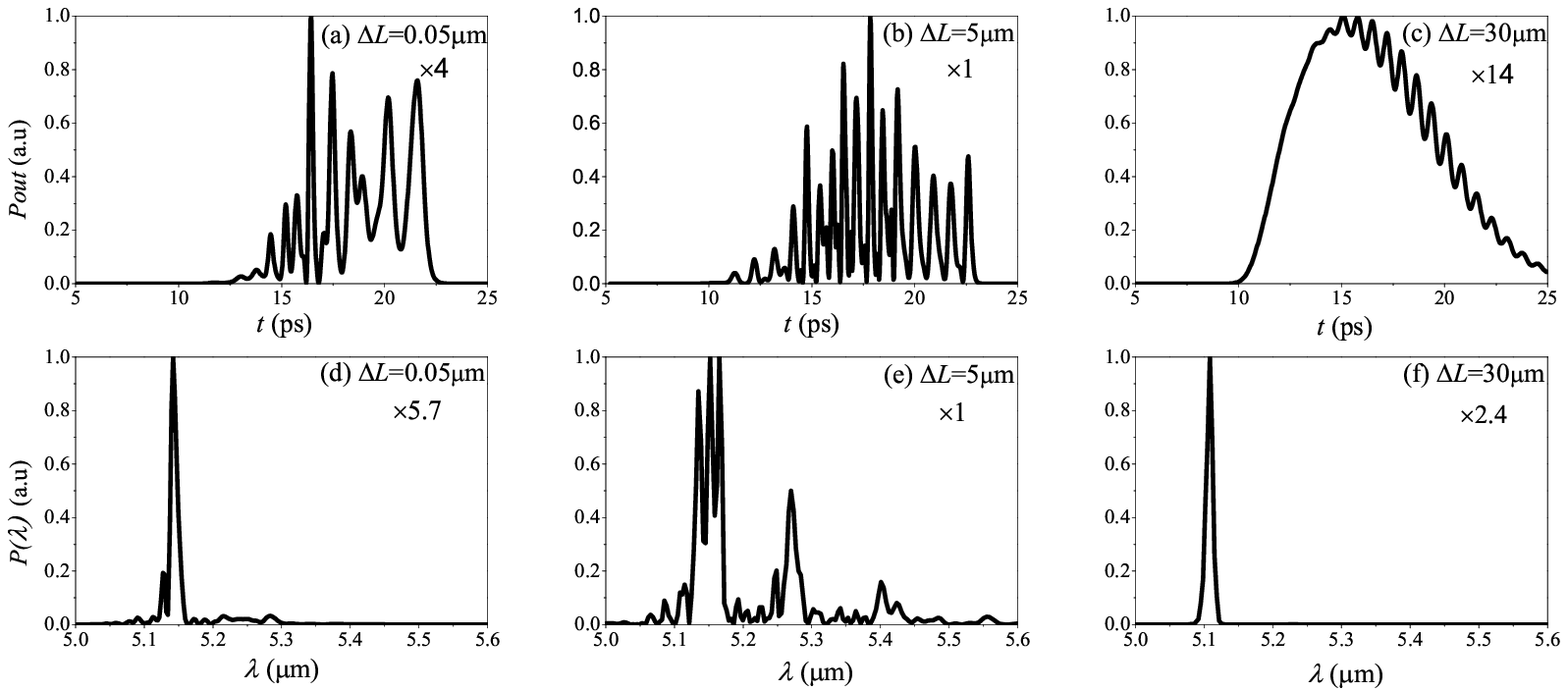}
\figcaption{\label{fig5}   The temporal and spectral structures of the FEL pulses at saturation with different detuning for the wavelength of 5 $\mu$m. Both the intensity values are normalized to those of $\Delta L=5$ $\mu$m.}
\end{center}
\begin{multicols}{2}
The temporal and spectral structures at saturation with different detuning for the wavelength of 5 $\mu$m are given in Fig. 5. The pulse RMS width can be tuned in the range of 3.2-6 ps with corresponding spectral width of 2-0.2\% when adjusting $\Delta L$ from 0.05 $\mu$m to 30 $\mu$m. At this time, a comb structure inside the optical pulse is generated, in which a single spike width is approximately comparable to the coherent length $l _c$ and the pulse width $\sigma _p$ is comparable to the electron bunch length $\sigma _e$ for optimal detuning. Furthermore, the pulse width dependence on $\Delta L$ have no obvious change in deep saturated region, such as Fig. 5 (a) and (b). Such behavior was predicted in Ref. \cite{lab24} and has been experimentally observed in several FELO groups \cite{lab25,lab26}.

\section{Analysis of the comb structure}

The super-mode theory and simulation results indicate that the normalized slippage length $\mu$ has a distinct influence on the output characteristics. After experiencing enough setting-up time for the optical field, the comb structure may emerge gradually.

Firstly, for 5 $\mu$m case, $\mu$ is equal to 0.2 ($\mu$$<$1). With the optimal cavity detuning length (small $\Delta L$), as shown in Fig. 6, the comb structure starts to emerge at the 50th roundtrip when the power is near saturated. The optical pulse duration also grows gradually with the increase of oscillation number. In deep saturated region, the width of a single spike is approximately comparable to the coherent length $\l _c$ which is calculated by Eq. (5) to be 130 $\mu$m. Moreover, from Fig. 5 (c), when $\Delta L = 30$ $\mu$m, the output power is small and the comb structure is less likely to occur. The unshown simulation results indicate that the comb structure will disappear when the cavity length detuning is larger. As $\Delta L$ decrease in some degree, the improved coupling between electrons and optical pulse generates higher power, and the comb structure inside the pulse appears. In this case, the comb structure is first appearing at the head of the optical pulse, then moving towards the tail. At small $\Delta L$, the output power is larger enough to make a rich comb structure occur.

\begin{center}
\includegraphics[height=5.5cm,width=8.5cm]{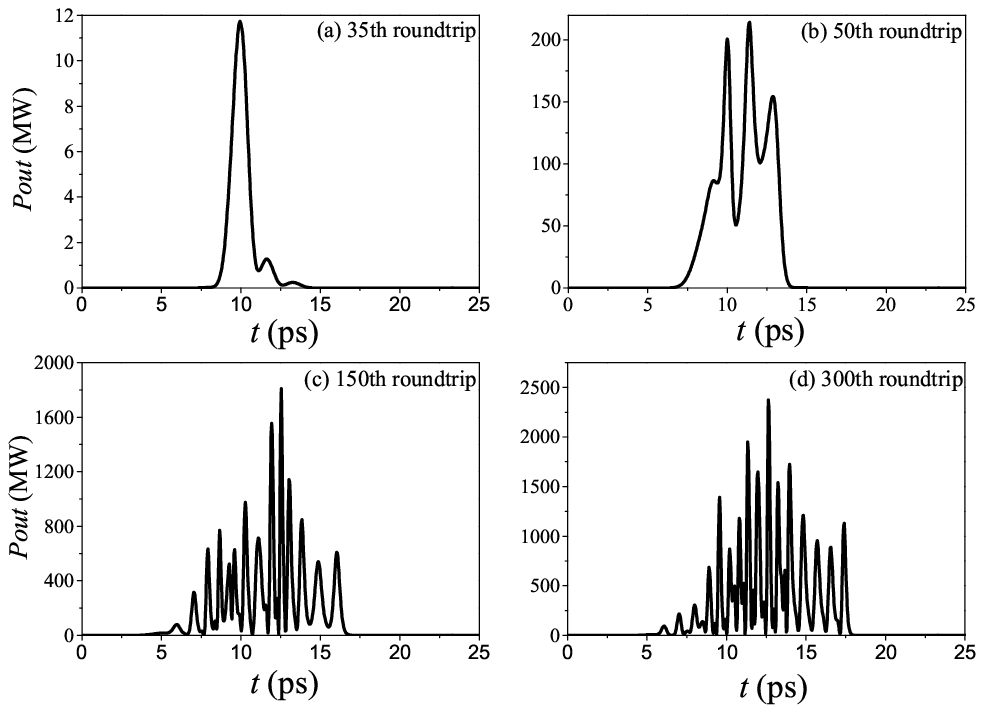}
\figcaption{\label{fig6}   The evolution of the temporal structure in different roundtrips for the wavelength of 5 $\mu$m. The cavity length detuning is 5 $\mu$m.}
\end{center}

Secondly, for 50 $\mu$m case, $\mu$ is equal to 2 ($\mu$$>$1). There is almost a single spike inside the optical pulse in Fig. 4. In order to further verify the influence of $\mu$ to the comb structure, we make $\sigma _e$ to be 2 and 4 times bigger than before and keep other parameters constant for 50 $\mu$m case, namely, $\mu$ is 2, 1 and 0.5. With the optimal detuning, the corresponding optical pulse structures are shown in Fig. 7.

The difference in 5 $\mu$m and 50 $\mu$m cases imply $\mu$ plays an important role in the emergence of comb structure. When $\mu$ is big enough, the comb structure phenomenon is not apparent; when $\mu$ becomes smaller, especially smaller than one, the comb structure phenomenon is remarkable in deep saturation region and each single spike width is comparable to $\l _c$. The corresponding width of optical pulse is comparable to $\sigma _e$. Furthermore, it should be taken notice of, however, that the FELO output power should be big enough to make the comb structure come out.

\begin{center}
\includegraphics[height=3cm,width=8.6cm]{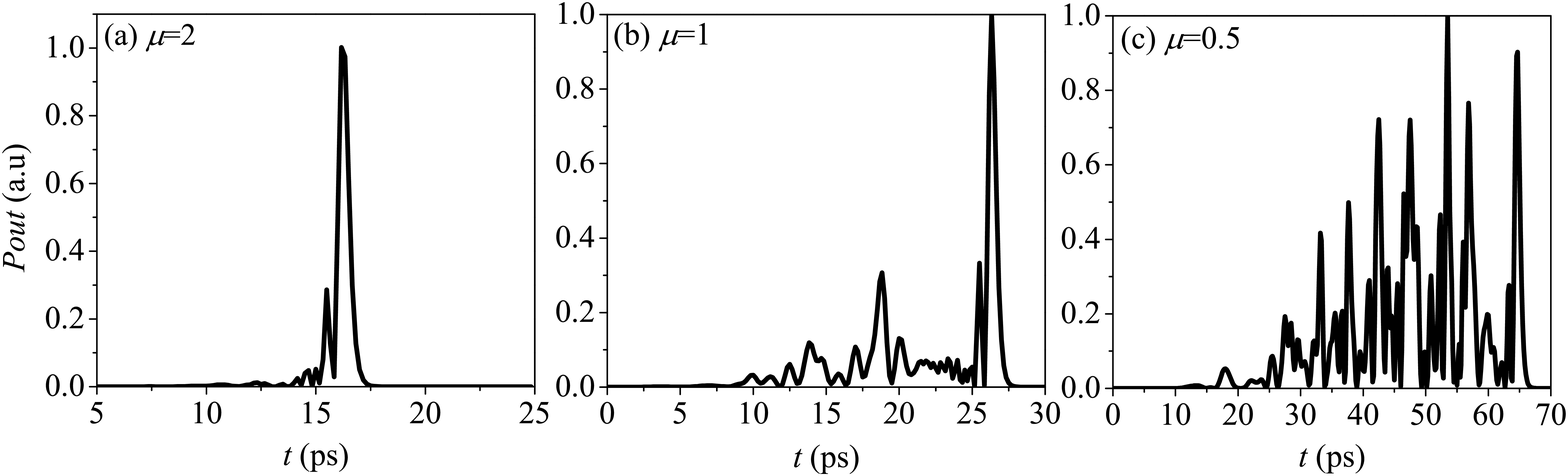}
\figcaption{\label{fig7}   The relationship between $\mu$ and comb structure inside the pulse in deeply saturated region for 50 $\mu$m case with the optimal detuning.}
\end{center}

\section{Conclusions}

In this paper, we have numerically studied the output characteristics of the mid-infrared FELO of FELiChEM with the cavity length detuning. The results show that the cavity length detuning has a significant impact on the power,  temporal and spectral structures of the output FEL pulse. The output pulse duration can be tuned in the range of 1-6 ps with corresponding bandwidth of 13-0.2\% by adjusting the cavity length detuning. Furthermore, the special discussion on the comb structure proves that it occurs in deep saturation region when the normalized slippage length is less than unity. The comb structure, i.e., mode-locked structure has been widely studied in high-gain FELs. This paper indicates the possibility of implementing natural mode-locking with conventional FEL oscillators. On the other hand, if not expected, it can be avoided by operating the oscillator with a relatively large detuning length. This work will provide a theoretical support for the operation of the mid-infrared FELO of FELiChEM and other similar facilities.

\end{multicols}

\vspace{-1mm}
\centerline{\rule{80mm}{0.1pt}}
\vspace{2mm}

\begin{multicols}{2}

\end{multicols}

\clearpage
\end{CJK*}
\end{document}